\documentclass{birkmult}
\usepackage{color}

\theoremstyle{definition}

\theoremstyle{remark}

\numberwithin{equation}{section}

\begin{document}
\title[New trends in the general relativistic Poynting-Robertson effect modeling]{New trends in the general relativistic Poynting-Robertson effect modeling}

\author[Vittorio De Falco]{Vittorio De Falco}

\address{Research Centre for Computational Physics and Data Processing, Faculty of Philosophy \& Science, Silesian University in Opava, Bezru\v{c}ovo n\'am.~13, CZ-746\,01 Opava, Czech Republic}

\email{vittorio.defalco@physics.cz}

\subjclass{XXXX; XXXX}

\keywords{XXXl}

\date{\today}

\begin{abstract}
The general relativistic Poynting-Robertson (PR) effect is a very important dissipative phenomenon occurring in high-energy astrophysics. Recently, it has been proposed a new model, which upgrades the two-dimensional (2D) description in the three-dimensional (3D) case in Kerr spacetime. The radiation field is considered as constituted by photons emitted from a rigidly rotating spherical source around the compact object. Such dynamical system admits the existence of a critical hypersurface, region where the gravitational and radiation forces balance and the matter reaches it at the end of its motion. Selected test particle orbits are displayed. We show how to prove the stability of these critical hypersurfaces within the Lyapunov theory. Then, we present how to study such effect under the Lagrangian formalism, explaining how to analytically derive the Rayleigh potential for the radiation force. In conclusion, further developments and future projects are discussed.
\end{abstract}

\maketitle
\section{Introduction}
\label{sec:intro}
The actual revolutionary discoveries occurred in the last four years represented by the detection of gravitational waves first from a binary black holes (BHs) \cite{Abott2016} and then from a neutron stars (NSs) \cite{Abott2017} systems and the first imaging of the matter motion around the supermassive BH in M87 Galaxy \cite{EHT2019} constitute a strong motivation to improve the actual theoretical models to validate Einstein theory or possible extension of it when benchmarked with the observations. The motion of relatively small-sized test particles, like dust grains or gas clouds, meteors, accretion disk matter elements, around radiating sources located outside massive compact objects is strongly affected by gravitational and radiation fields, and an important effect to be taken into account is the general relativistic PR effect \cite{Poynting1903,Robertson1937}. 

This phenomenon occurs each time the radiation field invests the test particle, raising up its temperature, which for the Stefan-Boltzmann law starts remitting radiation. This process of absorption and remission of radiation generates a recoil force opposite to the test body orbital motion. Such mechanism removes thus very efficiently angular momentum and energy from the test particle, forcing it to spiral inward or outward depending on the radiation field intensity. This effect has been extensively studied in Newtonian gravity within Classical Mechanics \cite{Poynting1903} and Special Relativity \cite{Robertson1937}, and then applied in the Solar system \cite{Burns1979}. Only in 2009 -- 2011, 
this model has been proposed in General Relativity (GR) by Bini and collaborators within the equatorial plane of the Ker spacetime \cite{Bini2009,Bini2011}. Recently, it has been extended in the 3D space in Kerr metric \cite{Defalco20193D,Bakala2019,Wielgus2019}. One of the most evident implications of such effect is the formation of stable structures, termed critical hypersurfaces, around the compact object \cite{Defalco2019ST}. This phenomenon has been analysed also under a Lagrangian formulation \cite{Defalco2018,Defalco2019EF1,Defalco2019EF2}. The novel aspects of such approach consists in the introduction of new techniques to deal with the non-linearities in gravity patterns based on two new fundamental aspects: (1) use of an integrating factor to make closed differential forms  \cite{Defalco2018}; (2) development of a new method termed \emph{energy formalism}, which permits to analytically determine the Rayleigh potential associated to the radiation force \cite{Defalco2019EF1,Defalco2019EF2}. 

The article is structured as follows: in Sec. \ref{sec:3D} the 3D model and its proprieties are described; in Sec. \ref{sec:stbch} the stability of the critical hypersurfaces is discussed within the Lyapunov theory; in Sec. \ref{sec:RF} we analytically determine the Rayleigh dissipation function by using the energy formalism. Finally in Sec. \ref{sec:end} the conclusions are drawn.

\section{General relativistic 3D PR effect model}
\label{sec:3D}
We consider a rotating compact object, whose geometry is described by the Kerr metric. Using the signature $(-,+,+,+)$ and geometrical units ($c = G = 1$), the metric line element, $ds^2=g_{\alpha\beta}dx^\alpha dx^\beta$, in Boyer-Lindquist coordinates, parameterized by mass $M$ and spin $a$, reads as \cite{Misner1973}
\begin{flushleft}
\begin{equation}\label{kerr_metric}
 \mathrm{d}s^2 = \left(\frac{2Mr}{\Sigma}-1\right)\mathrm{d}t^2 
  - \frac{4Mra\sin^2\theta}{\Sigma}\mathrm{d}t \mathrm{d}\varphi
+ \frac{\Sigma}{\Delta}\mathrm{d}r^2 
  + \Sigma\mathrm{d}\theta^2
  + \rho\sin^2\theta\mathrm{d}\varphi^2, 
\end{equation}
\end{flushleft}
where $\Sigma \equiv r^{2} + a^{2}\cos^{2}\theta$, $\Delta \equiv r^{2} - 2Mr + a^{2}$, and $\rho  \equiv r^2+a^2+2Ma^2r\sin^2\theta/\Sigma$. The determinant of the metric is $g=-\Sigma^2\sin^{2}\theta$. The orthonormal frame adapted to the zero angular momentum observers (ZAMOs) is \cite{Defalco20193D,Bakala2019}
\begin{equation} \label{eq:zamoframes}
\begin{aligned}
&\boldsymbol{e_{\hat t}}\equiv\boldsymbol{n}= \frac{(\boldsymbol{\partial_t}-N^{\varphi}\boldsymbol{\partial_\varphi})}{N},\quad
\boldsymbol{e_{\hat r}}=\frac{\boldsymbol{\partial_r}}{\sqrt{g_{rr}}},\quad
\boldsymbol{e_{\hat \theta}}=\frac{\boldsymbol{\partial_\theta}}{\sqrt{g_{\theta \theta }}},\quad
\boldsymbol{e_{\hat \varphi}}=\frac{\boldsymbol{\partial_\varphi}}{\sqrt{g_{\varphi \varphi }}}.
\end{aligned}
\end{equation}
where $N=(-g^{tt})^{-1/2}$ and $N^{\varphi}=g_{t\varphi}/g_{\varphi\varphi}$. The nonzero ZAMO kinematical quantities in the decomposition of the ZAMO congruence are acceleration $\boldsymbol{a}(n)=\nabla_{\boldsymbol{n}} \boldsymbol{n}$, expansion tensor along the $\hat{\varphi}$-direction $\boldsymbol{\theta_{\hat\varphi}}(n)$, and the relative Lie curvature vector $\boldsymbol{k_{(\rm Lie)}}(n)$ (see Table 1 in \cite{Defalco20193D}, for their explicit expressions).
 
The radiation field is modeled as a coherent flux of photons traveling along null geodesics on the Kerr metric. The related stress-energy tensor is \cite{Defalco20193D,Bakala2019}
\begin{equation}\label{STE}
T^{\mu\nu}=\mathcal{I}^2 k^\mu k^\nu\,,\qquad k^\mu k_\mu=0,\qquad k^\mu \nabla_\mu k^\nu=0,
\end{equation}
where $\mathcal{I}$ is a parameter linked to the radiation field intensity and $\boldsymbol{k}$ is the photon four-momentum field. Splitting $\boldsymbol{k}$ with respect to the ZAMO frame, we obtain \cite{Bakala2019}
\begin{eqnarray}
&&\boldsymbol{k}=E(n)[\boldsymbol{n}+\boldsymbol{\hat{\nu}}(k,n)], \label{photon1}\\
&&\boldsymbol{\hat{\nu}}(k,n)=\sin\zeta\sin\beta\ \boldsymbol{e_{\hat r}}+\cos\zeta\ \boldsymbol{e_{\hat\theta}}+\sin\zeta \cos\beta\ \boldsymbol{e_{\hat\varphi}}, \label{photon2}
\end{eqnarray}
where $E(n)$ is the photon energy measured in the ZAMO frame, $\boldsymbol{\hat{\nu}}(k,n)$ is the photon spatial unit relative velocity with respect to the ZAMOs, $\beta$ and $\zeta$ are the two angles measured in the ZAMO frame in the azimuthal and polar direction, respectively. The radiation field is governed by the two impact parameters $(b,q)$, associated respectively with the two emission angles $(\beta,\zeta)$. The radiation field photons are emitted from a spherical rigid surface having a radius $R_\star$ centered at the origin of the Boyer-Lindquist coordinates, and rotating with angular velocity $\Omega_{\mathrm{\star}}$. The photon impact parameters are \cite{Bakala2019}
\begin{eqnarray} 
&&b=-\left[\frac{\mathrm{g_{t\varphi}}+\mathrm{g_{\varphi\varphi}}\Omega_{\star} }{\mathrm{g_{tt}}+\mathrm{g_{t\varphi}} \Omega_{\star}}\right]_{r=R_\star},\label{kerr_impact_parameter}\quad
q=\left[b^{2}\cot^{2} \theta-a^{2} \cos^{2}\theta \right]_{r=R_\star}. \label{q_r}
\end{eqnarray}
The related photon angles in the ZAMO frame are \cite{Bakala2019}
\begin{equation} \label{ANG1}
\cos\beta=\frac{b N}{\sqrt{g_{\varphi\varphi}}(1+b N^\varphi)}, \qquad \zeta=\pi/2. 
\end{equation}
The parameter $\mathcal{I}$ has the following expression \cite{Bakala2019}
\begin{equation}\label{INT_PAR}
\mathcal{I}^2=\frac{\mathcal{I}_0^2}{\sqrt{\left( r^{2} + a^{2}-ab \right)^{2}- \Delta \left[ q + \left( b - a \right) ^{2} \right]}},
\end{equation}
where $\mathcal{I}_0$ is $\mathcal{I}$ evaluated at the emitting surface.

A test particle moves with a timelike four-velocity $\boldsymbol{U}$ and a spatial three-velocity with respect to the ZAMO frames, $\boldsymbol{\nu}(U,n)$, which both read as \cite{Bakala2019}
\begin{eqnarray} 
&&\boldsymbol{U}=\gamma(U,n)[\boldsymbol{n}+\boldsymbol{\nu}(U,n)], \label{testp}\\
&&\boldsymbol{\nu}=\nu(\sin\psi\sin\alpha\boldsymbol{e_{\hat r}}+\cos\psi\boldsymbol{e_{\hat\theta}}+\sin\psi \cos\alpha \boldsymbol{e_{\hat\varphi}}),
\end{eqnarray}
where $\gamma(U,n)\equiv\gamma=1/\sqrt{1-||\boldsymbol{\nu}(U,n)||^2}$ is the Lorentz factor, $\nu=||\boldsymbol{\nu}(U,n)||$, $\gamma(U,n) =\gamma$. We have that $\nu$ represents the magnitude of the test particle spatial velocity $\boldsymbol{\nu}(U,n)$, $\alpha$ is the azimuthal angle of the vector $\boldsymbol{\nu}(U,n)$ measured clockwise from the positive $\hat\varphi$ direction in the $\hat{r}-\hat{\varphi}$ tangent plane in the ZAMO frame, and $\psi$ is the polar angle of the vector $\boldsymbol{\nu}(U,n)$ measured from the axis orthogonal to the $\hat{r}-\hat{\varphi}$ tangent plane in the ZAMO frame. 

We assume that the radiation test particle interaction occurs through Thomson scattering, characterized by a constant momentum-transfer cross section $\sigma$, independent from direction and frequency of the radiation field. We can split the photon four momentum (\ref{photon1}) in terms of the velocity $\boldsymbol{U}$ as \cite{Bakala2019}
\begin{equation}
\boldsymbol{k}=E(U)[\boldsymbol{U}+\boldsymbol{\hat{\mathcal{V}}}(k,U)],
\end{equation}
where $E(U)$ is the photon energy measured by the test particle. The radiation force can be written as \cite{Bakala2019}
\begin{equation} \label{radforce}
{\mathcal F}_{\rm (rad)}(U)^{\hat \alpha}\equiv-\tilde{\sigma\mathcal{I}}^2(T^{\hat \alpha}{}_{\hat \beta} U^{\hat \beta}+U^{\hat \alpha} T^{\hat \mu}{}_{\hat \beta} U_{\hat \mu} U^{\hat \beta})=\tilde{\sigma} \, [\mathcal{I} E(U)]^2\, \hat{\mathcal V}(k,U)^{\hat \alpha},
\end{equation}
where $m$ is the test particle mass and the term $\tilde{\sigma}[\mathcal{I} E(U)]^2$ reads as \cite{Bakala2019} 
\begin{equation} \label{eq: sigma_tilde}
\tilde{\sigma}[\mathcal{I} E(U)]^2=\frac{ A\,\gamma^2(1+b N^\varphi)^2[1-\nu\sin\psi\cos(\alpha-\beta)]^2}{N^2\sqrt{\left( r^{2} + a^{2}-ab \right)^{2}- \Delta \left[ q + \left( b - a \right) ^{2} \right]}},
\end{equation}
with $A=\tilde{\sigma}[\mathcal{I}_0 E_p]^2$ being the luminosity parameter, which can be equivalently written as $A/M=L/L_{\rm EDD}\in[0,1]$ with $L$ the emitted luminosity at infinity and $L_{\rm EDD}$ the Eddington luminosity, and $E_p=-k_t$ is the conserved photon energy along the test particle trajectory. The terms $\hat{\mathcal V}(k,U)^{\hat \alpha}$ are the radiation field components, whose expressions are \cite{Bakala2019}
\begin{eqnarray}\label{rad}
&&\hat{\mathcal{V}}^{\hat r}=\frac{\sin\beta}{\gamma [1-\nu\sin\psi\cos(\alpha-\beta)]}-\gamma\nu\sin\psi\sin\alpha, \quad \hat{\mathcal{V}}^{\hat \theta}=-\gamma\nu\cos\psi,\\
&&\hat{\mathcal{V}}^{\hat\varphi}=\frac{\cos\beta}{\gamma [1-\nu\sin\psi\cos(\alpha-\beta)]}-\gamma\nu\sin\psi\cos\alpha,\quad
\hat{\mathcal{V}}^{\hat t}=\gamma\nu\left[\frac{\sin\psi\cos(\alpha-\beta)-\nu}{1-\nu\sin\psi\cos(\alpha-\beta)}\right].\notag
\end{eqnarray}

Collecting all the information together, it is possible to derive the resulting equations of motion for a test particle moving in a 3D space, which are \cite{Bakala2019}
\begin{eqnarray}
&&\frac{d\nu}{d\tau}= -\frac{1}{\gamma}\left\{ \sin\alpha \sin\psi\left[a(n)^{\hat r}\right.+2\nu\cos \alpha\sin\psi\, \theta(n)^{\hat r}{}_{\hat \varphi} \right]\label{EoM1}\\
&&\left.\ \quad +\cos\psi\left[a(n)^{\hat \theta}+2\nu\cos\alpha\sin\psi\, \theta(n)^{\hat \theta}{}_{\hat \varphi}\right]\right\}+\frac{\tilde{\sigma}[\Phi E(U)]^2}{\gamma^3\nu}\hat{\mathcal{V}}^{\hat t},\nonumber\\
&&\frac{d\psi}{d\tau}= \frac{\gamma}{\nu} \left\{\sin\psi\left[a(n)^{\hat \theta}+k_{\rm (Lie)}(n)^{\hat \theta}\,\nu^2 \cos^2\alpha+2\nu\cos \alpha\sin^2\psi\ \theta(n)^{\hat \theta}{}_{\hat \varphi}\right]\right.\nonumber\\
&&\left.\ \quad-\sin \alpha\cos\psi \left[a(n)^{\hat r}+k_{\rm (Lie)}(n)^{\hat r}\,\nu^2+2\nu\cos \alpha\sin\psi\, \theta(n)^{\hat r}{}_{\hat \varphi}\right]\right\}\label{EoM2}\\
&&\ \quad+\frac{\tilde{\sigma}[\Phi E(U)]^2}{\gamma\nu^2\sin\psi}\left[\hat{\mathcal{V}}^{\hat t}\cos\psi-\hat{\mathcal{V}}^{\hat \theta}\nu\right],\nonumber
\end{eqnarray}
\begin{eqnarray}
&&\frac{d\alpha}{d\tau}=-\frac{\gamma\cos\alpha}{\nu\sin\psi}\left[a(n)^{\hat r}+2\theta(n)^{\hat r}{}_{\hat \varphi}\ \nu\cos\alpha\sin\psi+k_{\rm (Lie)}(n)^{\hat r}\,\nu^2\right.\label{EoM3}\\
&&\left.\ \quad+k_{\rm (Lie)}(n)^{\hat \theta}\,\nu^2\cos^2\psi \sin\alpha\right]+\frac{\tilde{\sigma}[\Phi E(U)]^2\cos\alpha}{\gamma\nu\sin\psi}\left[\hat{\mathcal{V}}^{\hat r}-\hat{\mathcal{V}}^{\hat \varphi}\tan\alpha\right],\nonumber\\
&&U^{\hat r}\equiv\frac{dr}{d\tau}=\frac{\gamma\nu\sin\alpha\sin\psi}{\sqrt{g_{rr}}}, \label{EoM4}\\
&&U^{\hat \theta}\equiv\frac{d\theta}{d\tau}=\frac{\gamma\nu\cos\psi}{\sqrt{g_{\theta\theta}}} \label{EoM5},\\
&&U^{\hat \varphi}\equiv\frac{d\varphi}{d\tau}=\frac{\gamma\nu\cos\alpha\sin\psi}{\sqrt{g_{\varphi\varphi}}}-\frac{\gamma N^\varphi}{N},\label{EoM6}\\
&&U^{\hat t}\equiv \frac{dt}{d\tau}=\frac{\gamma}{N},\label{time}
\end{eqnarray}
where $\tau$ is the affine parameter along the test particle trajectory. 

\subsection{Critical hypersurfaces}
\label{sec:critc_rad}
\begin{figure}[t!]
	\centering
	\includegraphics[scale=0.64]{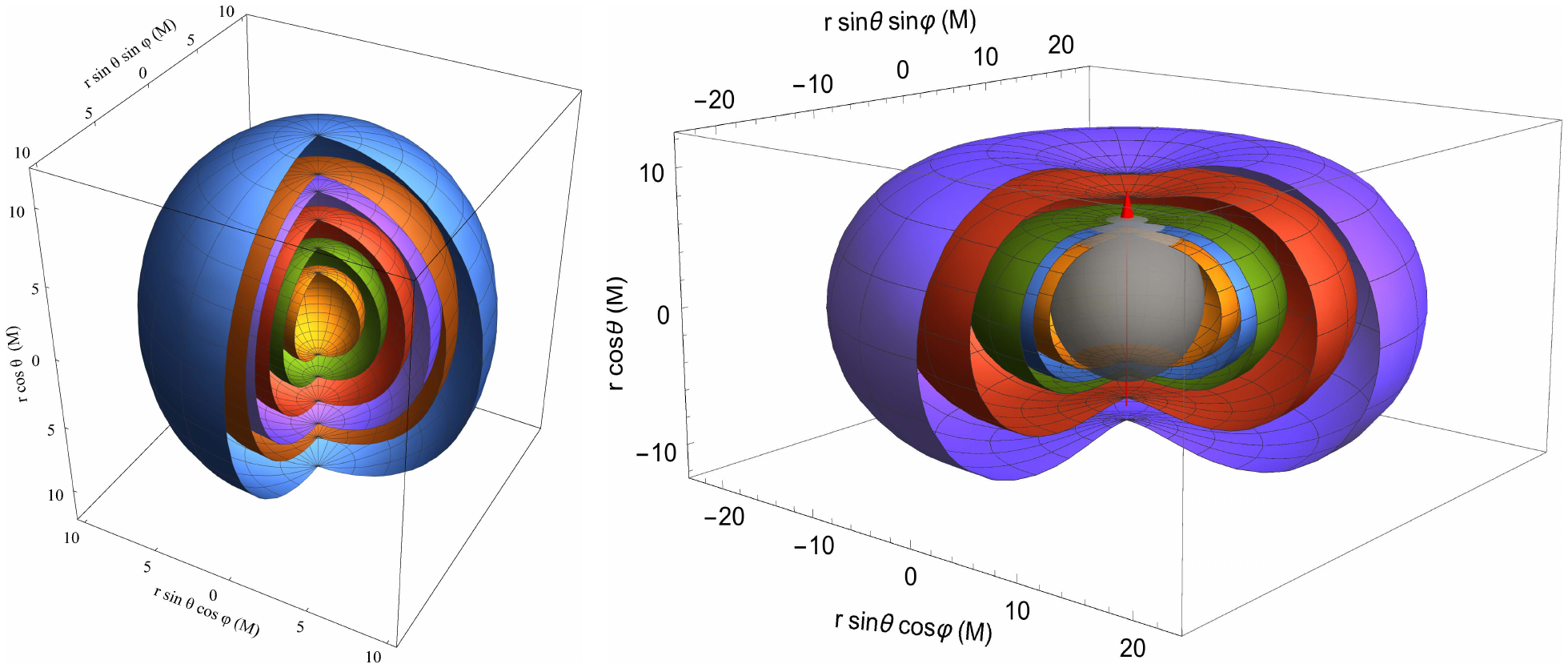}
	\caption{Left panel: Critical hypersurfaces for $\Omega_\star=0$ and the luminosity parameters $A=0.5, \,0.7, \,0.8, \,0.85, \,0.87, \,0.9$ at a constant spin $a=0.9995$. The respective critical radii in the equatorial plane are $r^{\rm eq}_{\rm(crit)} \sim 2.71M,  4.01M, 5.52M, 7.04M, 7.99M, 10.16M$, while at poles they are $r^{\rm pole}_{\rm(crit)} \sim 2.97M,  4.65M,  6.56M,  8.38M,  9.48M, 11.9M$.
	Right panel: Critical hypersurfaces for a NS (grey sphere) with $\Omega_\star=0.031$, $R_\star=6M$, and luminosity parameters $A=0.75,\, 0.78, \,0.8, \,0.85, \,0.88$ at a constant spin $a=0.41$. The respective critical radii in the equatorial plane are $r^{\rm eq}_{\rm(crit)} \sim 8.88M,\  10.61M,\ 12.05M,\ 17.26M,\ 22.43M,\ $, while at poles they are $r^{\rm pole}_{\rm(crit)} \sim 4.73M,\  5.28M,\  5.74M,\  7.43M,\  9.11M$. The red arrow is the polar axis.}
	\label{fig:Fig1}
\end{figure}

The dynamical system defined by Eqs. (\ref{EoM1})--(\ref{EoM6}) exhibits a critical hypersurface outside around the compact object, where there exists a balance among gravitational and radiation forces, see Fig. \ref{fig:Fig1}. 
On such region the test particle moves purely circular with constant velocity ($\nu=\mbox{const}$) with respect to the ZAMO frame ($\alpha=0,\pi$), and the polar axis orthogonal to the critical hypersurface ($\psi=\pm\pi/2$). These requirements entail $d\nu/d\tau=d\alpha/d\tau=0$, from which we have \cite{Bakala2019}
\begin{eqnarray}
&&\nu=\cos\beta, \label{eq:crit_hyper1} \\
&&a(n)^{\hat r}+2\theta(n)^{\hat r}{}_{\hat\varphi}\nu+k_{\rm (Lie)}(n)^{\rm \hat r}\nu^2\label{eq:crit_hyper2}\\
&&=\frac{A(1+bN^\varphi)^2\sin^3\beta}{N^2\gamma\sqrt{\left( r_{\rm (crit)}^{2} + a^{2}-ab \right)^{2}- \Delta_{\rm (crit)} \left[ q + \left( b - a \right) ^{2} \right]}},\notag 
\end{eqnarray}
where the first condition means that the test particle moves on the critical hypersurface with constant velocity equal to the azimuthal photon velocity; whereas the second condition determine the critical radius $r_{\rm (crit)}$ as a function of the polar angle through an implicit equation, once $A,a,R_\star,\Omega_\star$ are assigned. 

\begin{figure}[t!]
	\centering
	\hbox{
	\includegraphics[scale=0.25]{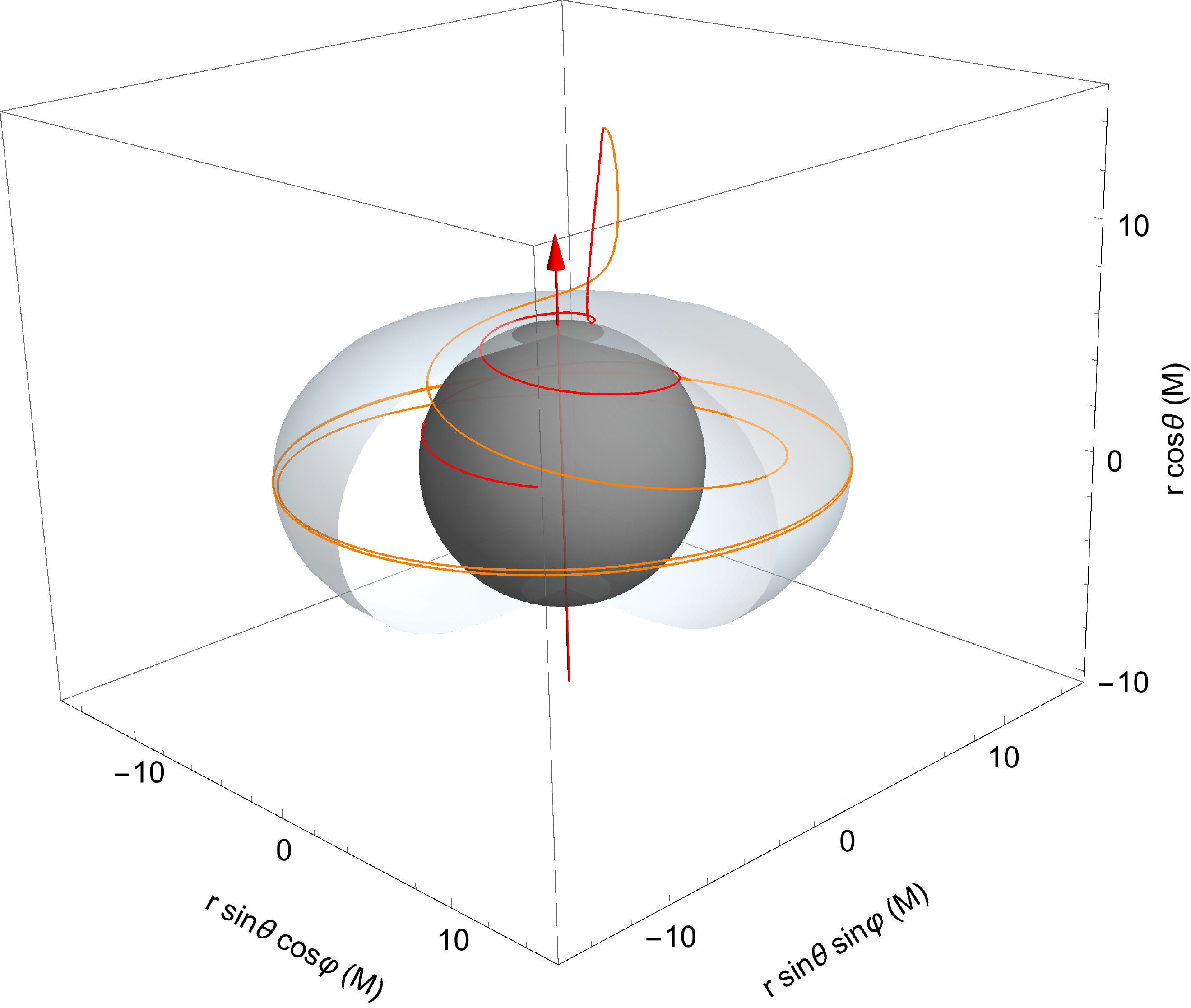}
	\hspace{0.4cm}
	\includegraphics[scale=0.2]{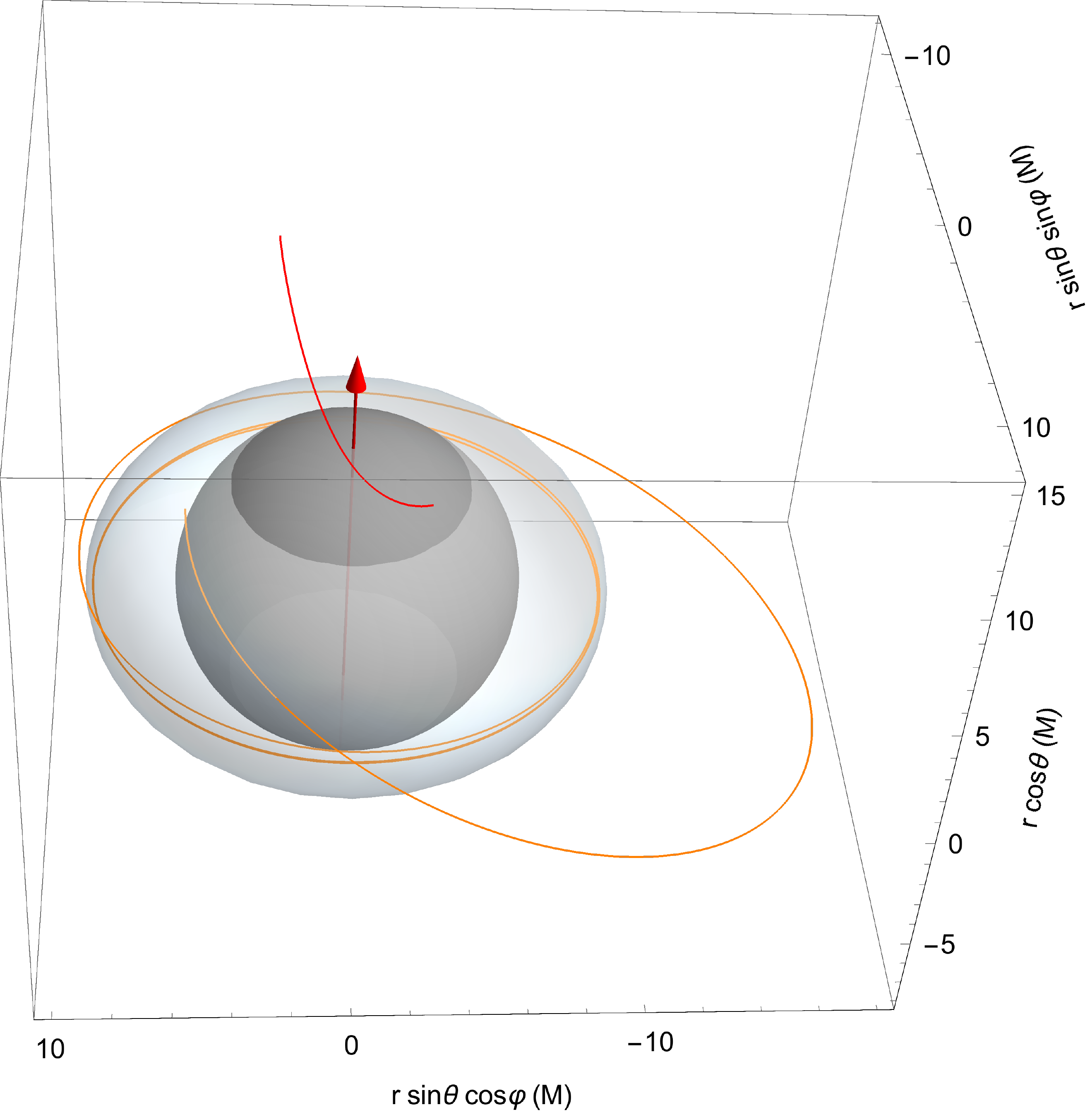}}
	\caption{Left panel: Test particle trajectories around a NS of spin $a=0.41$, radius $R_\star=6M$, angular velocity $\Omega_\star=0.031$, and luminosity parameter $A=0.8$, starting at the position $(r_0,\theta_0)=(15M,10^\circ)$ with the initial velocity $\nu_0=0.01$ oriented in the azimuthal corotating direction direction (orange) and oriented radially towards the emitting surface (red).
Right panel:Test particle trajectories around a NS of spin $a=0.07$, radius $R_\star=6M$, angular velocity $\Omega_\star=0.005$, and luminosity parameter $A=0.85$, starting at the position $(r_0,\theta_0)=(15M,10^\circ)$ with the initial velocity $\nu_0=0.01$ oriented in the azimuthal corotating direction direction (orange) and oriented radially towards the emitting surface (red). The black sphere corresponds to the emitting surface of the NS. The blue-gray surface denotes the critical hypersurface.}
	\label{fig:Fig2}
\end{figure}

In general we have  $d\psi/d\tau\neq0$, because the $\psi$ angle change during the test particle motion on the critical hypersurface, having the so-called \emph{latitudinal drift}. This effect, occurring for the interplay of gravitational and radiation actions in the polar direction, brings definitively the test particle on the equatorial plane \cite{Defalco20193D,Bakala2019}. Only for $\psi=\theta=\pi/2$, we have $d\psi/d\tau=0$, corresponding to the equatorial ring. However, we can have $d\psi/d\tau=0$, also for a $\theta=\bar{\theta}\neq\pi/2$, having the so-called \emph{suspended orbits}. The condition for this last configuration for $b\neq0$ reads as \cite{Bakala2019}
\begin{equation}
\begin{aligned}\label{eq:susporbit}
&a(n)^{\hat \theta}+k_{\rm (Lie)}(n)^{\hat \theta}\,\nu^2+2\nu\sin^2\psi\ \theta(n)^{\hat \theta}{}_{\hat \varphi}\\
&+\frac{A (1+bN^\varphi)^2(1-\cos^2\beta\sin\psi)\cos\beta}{\gamma N^2\sqrt{\left( r_{\rm (crit)}^{2} + a^{2}-ab \right)^{2}- \Delta_{\rm (crit)} \left[ q + \left( b - a \right) ^{2} \right]}\tan\psi}=0,
\end{aligned}
\end{equation}
which permits to be solved in terms of $\psi$. Instead for $b=0$ we obtain $\psi=\pm\pi/2$ \cite{Defalco20193D}. The critical points are either the suspended orbits or the equatorial ring, where the test particle ends its motion. In Fig. \ref{fig:Fig2} we display some selected test particle trajectories to give an idea how the PR effect alters the matter motion surrounding a radiation source around a compact object \cite{Bakala2019}. 

\section{Stability of the critical hypersurfaces}
\label{sec:stbch}
To prove the stability of the critical hypersurfaces, we consider only those initial configurations, where the test particle ends its motion on them without escaping at infinity. Once the stability has been proven, it immediately follows that the critical equatorial ring is a stable attractor (region where the test particle is attracted for ending its motion), and the whole critical hypersurface is a basin of attraction \cite{Defalco2019ST}.  

Bini and collaborators have proved it only in the Schwarzschild case within the linear stability theory (see Appendix in Ref. \cite{Bini2011}). This method consists in linearizing the dynamical system towards the critical points of the critical hypersurface and then looking at its eigenvalues. Theoretically such method is simple, but practically it implies several calculations (especially in the Kerr case).

There is a simpler, and more physical approach based on the Lyapunov theory. The dynamical system (\ref{EoM1})--(\ref{EoM6}), $\dot{\boldsymbol{x}}=\boldsymbol{f}(\boldsymbol{x})$, is defined in the domain $\mathcal{D}$, while the critical hypersurface is defined by $\mathcal{H}$. Let $\Lambda=\Lambda(\boldsymbol{x})$ be a real valued function, continuously differentiable in all points of $\mathcal{D}$, then $\Lambda$ is a Lyapunov function for $\dot{\boldsymbol{x}}=\boldsymbol{f}(\boldsymbol{x})$ if it fulfills the following conditions:
\begin{eqnarray}  
{\rm (I)}&&\quad \Lambda(\boldsymbol{x})>0,\quad \forall \boldsymbol{x}\in \mathcal{D}\setminus\mathcal{H};\label{eq:lia1}\\
{\rm (II)}&&\quad \Lambda(\boldsymbol{x_0})=0,\quad \forall \boldsymbol{x_0}\in \mathcal{H};\label{eq:lia2}\\ 
{\rm (III)}&&\quad \dot{\Lambda}(\boldsymbol{x})\equiv\nabla\Lambda(\boldsymbol{x})\cdot \boldsymbol{f}(\boldsymbol{x})\le0 ,\quad \forall \boldsymbol{x}\in \mathcal{D}\label{eq:lia3}.
\end{eqnarray}
Once the Lyapunov function $\Lambda$ has been found for all points belonging to the critical hypersurface $\mathcal{H}$, a theorem due to Lyapunov assures that $\mathcal{H}$ is stable \cite{Defalco2019ST}. 

The advantage to use this approach relies on easily studying the behavior of a dynamical system without knowing the analytical solution. The Lyapunov function is not unique and there is no fixed rules to determine it, indeed several times one is guided by the physical intuitions. For the general relativistic PR effect three different Lyapunov functions have been determined. The proof that they are Lyapunov function is based on expanding all the kinematic terms with respect to the radius estimating thus their magnitude (see Ref. \cite{Defalco2019ST}, for further details).

\begin{figure}[h!]
	\centering
	\vbox{
	\hbox{\hspace{0cm}
		\includegraphics[scale=0.25]{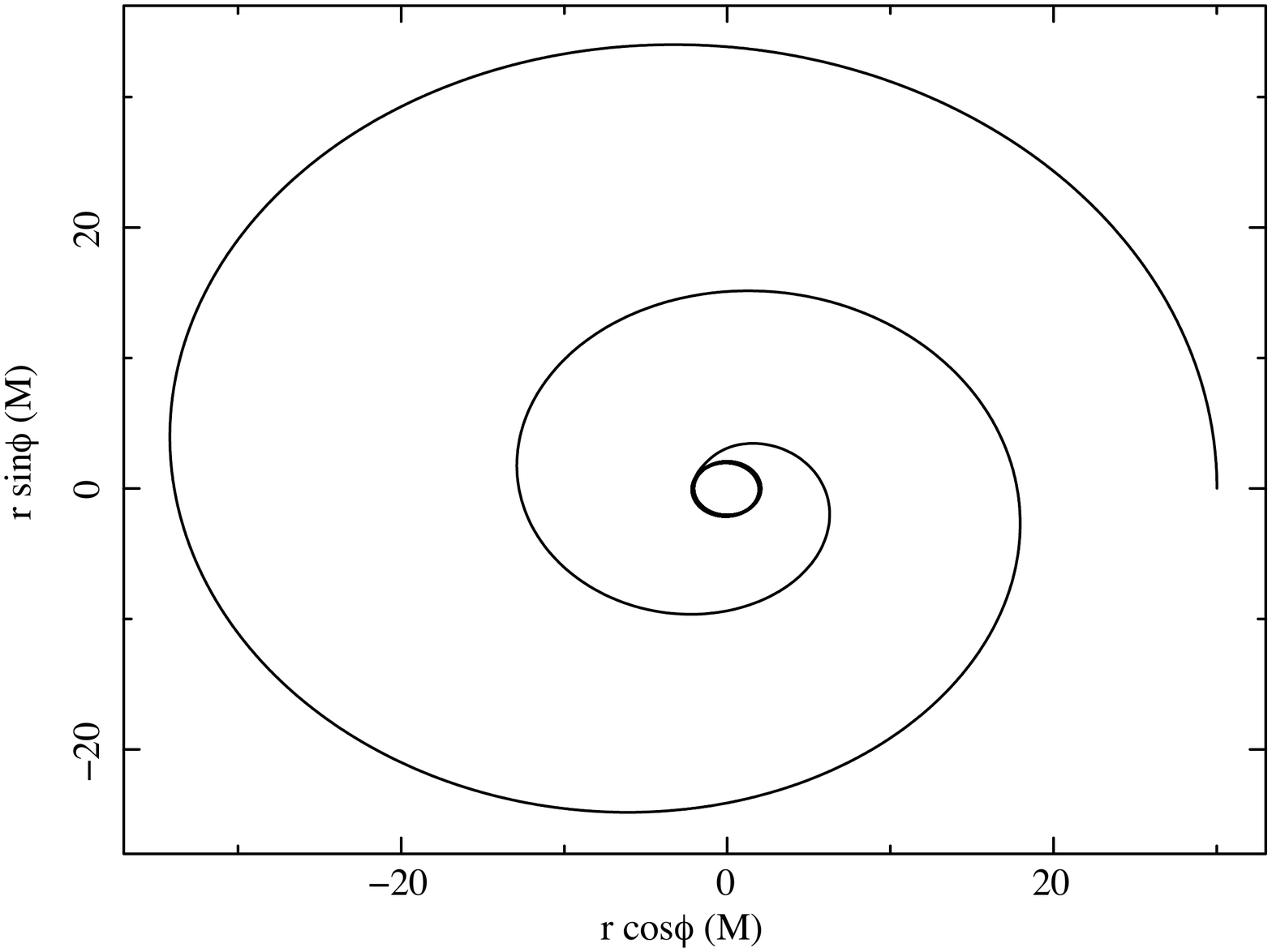}
		\hspace{-1 cm}
		\includegraphics[scale=0.25]{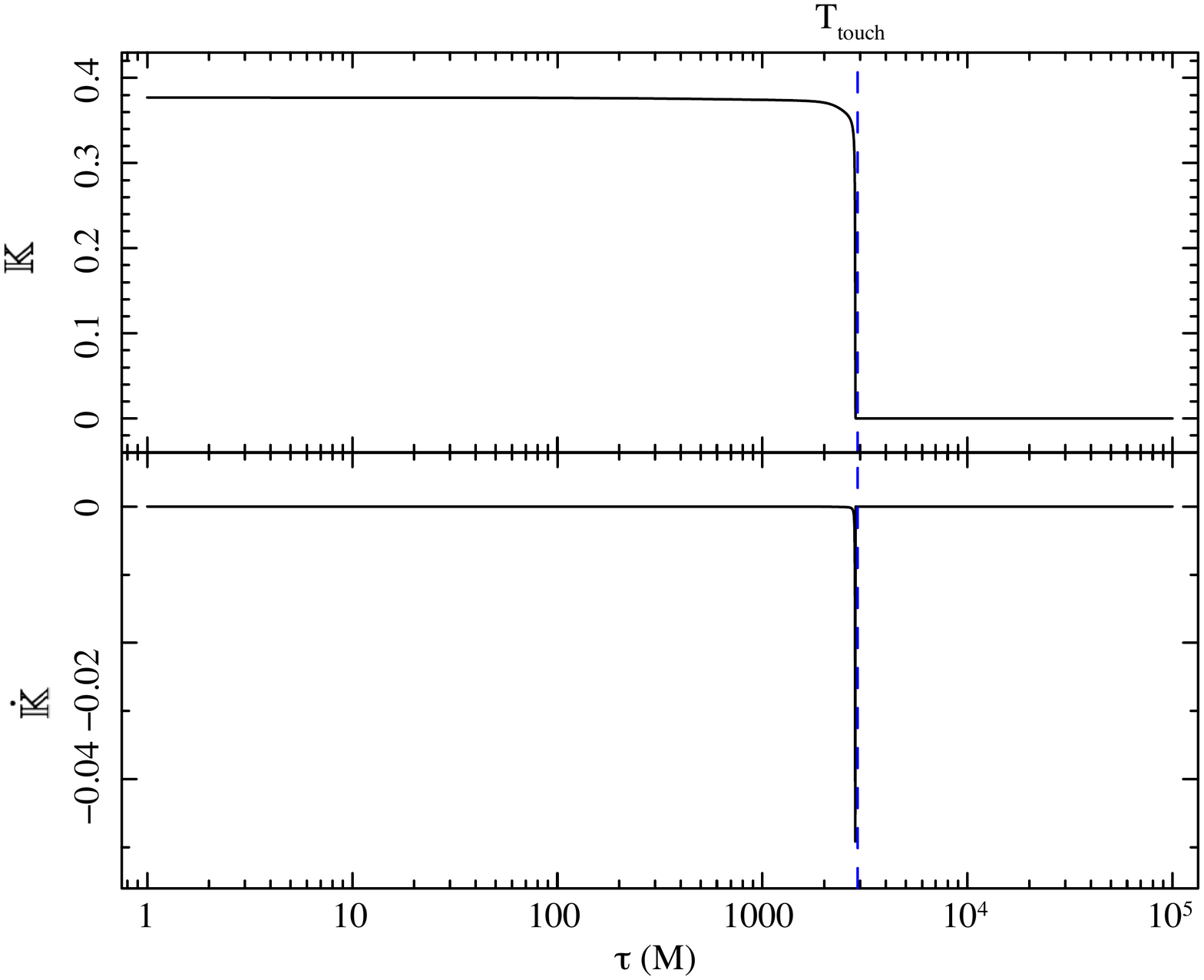}}
		\vspace{-0.3 cm}
	\hbox{\hspace{0cm}
		\includegraphics[scale=0.25]{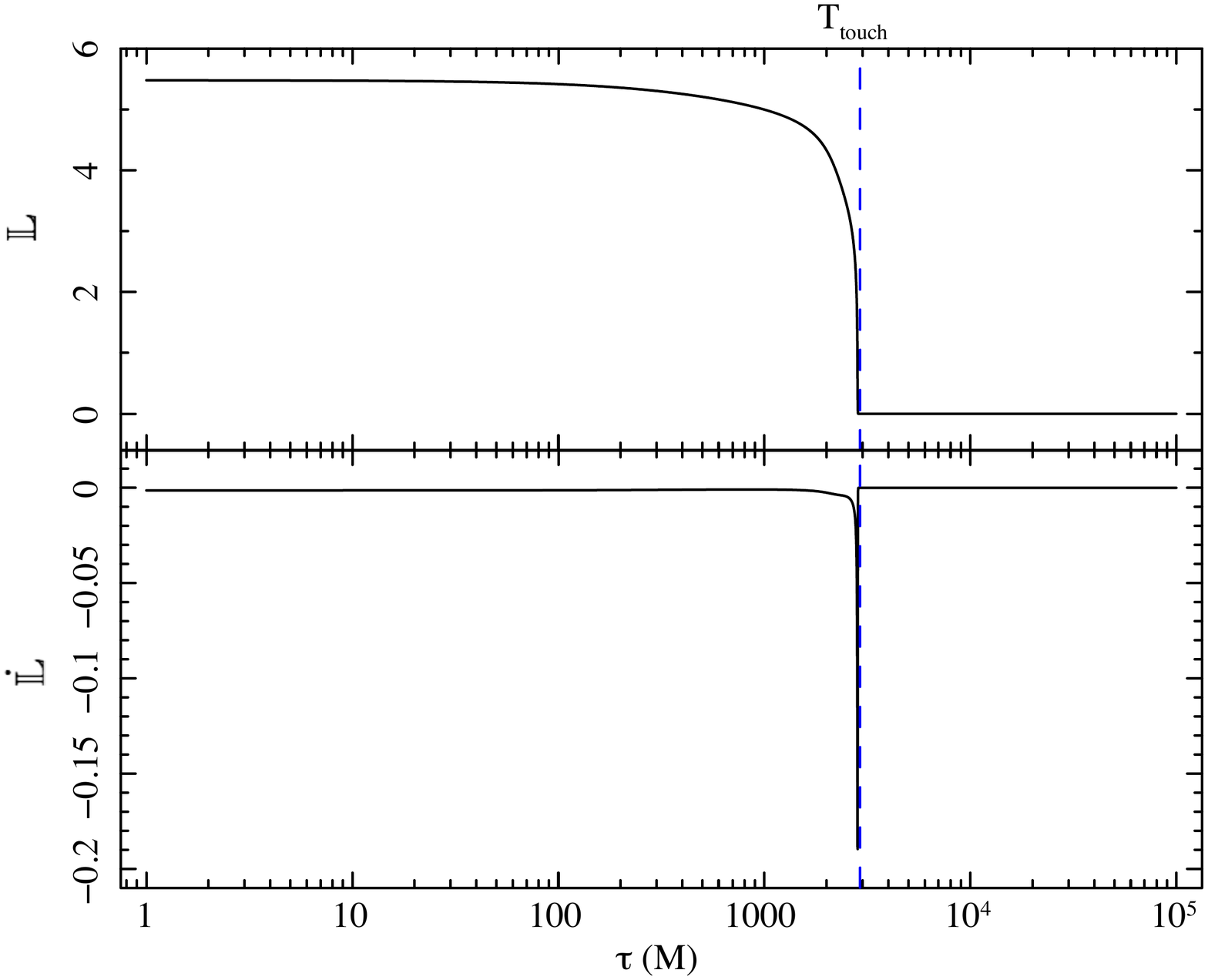}
		\hspace{-1 cm}
		\includegraphics[scale=0.25]{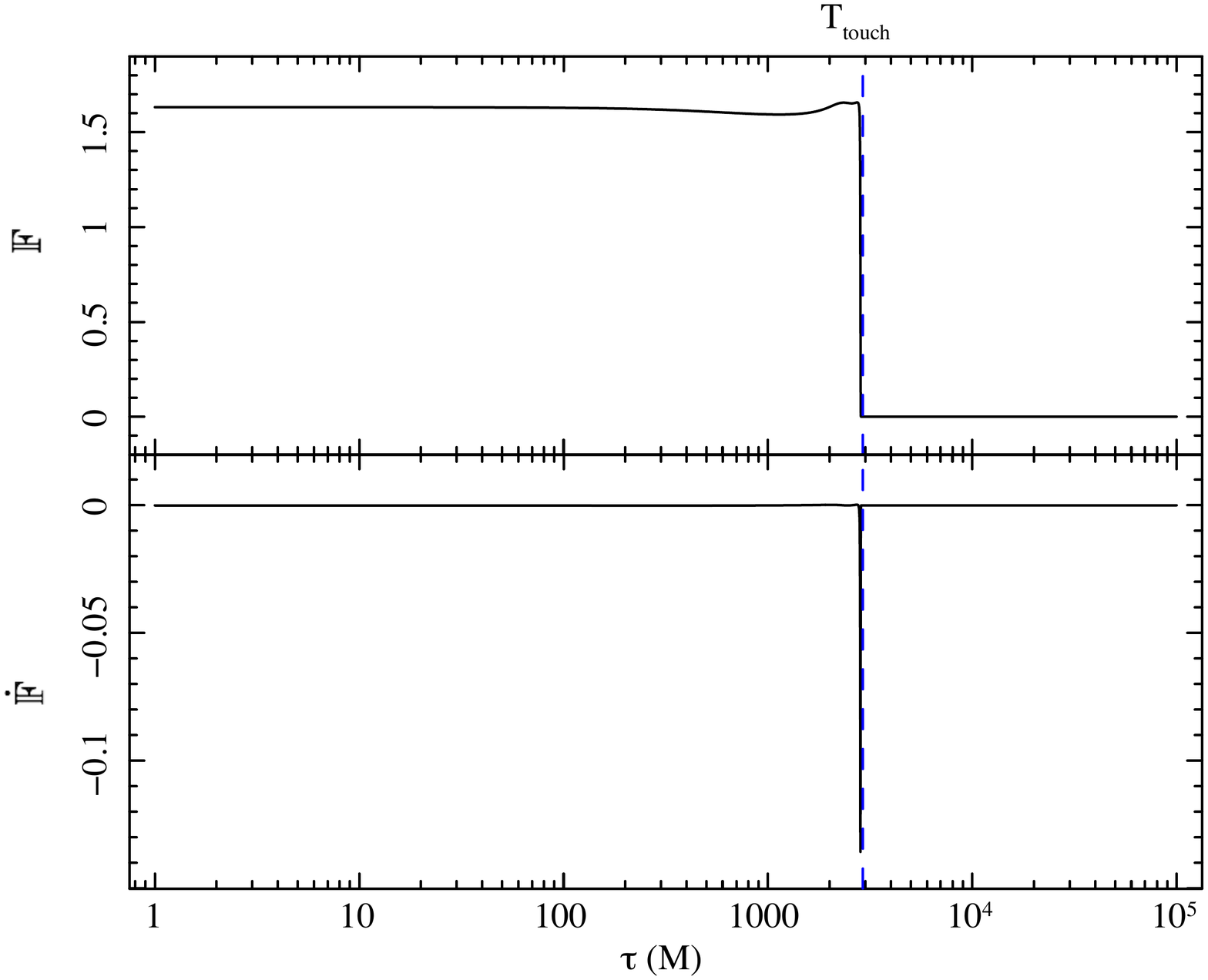}}}
	\caption{We show a test particle orbit and the related three Lyapunov functions. \emph{Upper left panel:} test particle moving around a rotating compact object with mass $M=1$, spin $a=0.3$, luminosity parameter $A=0.2$, and photon impact parameter $b=0$. The test particle starts its motion at the position $(r_0,\varphi_0)=(30M,0)$ with velocity $(\nu_0,\alpha_0)=(\sqrt{M/r_0},0)$. The critical hypersurface is a circle with radius $r_{\rm (crit)}=2.07M$. The energy (see Eqs. (\ref{eq:LF1}) and (\ref{eq:DLF1}), and \emph{upper right panel}), the angular momentum (see Eqs. (\ref{eq:LF2}) and (\ref{eq:DLF2}), and \emph{lower left panel}), and the Rayleigh potential (see Eqs. (\ref{eq:LF3}) and (\ref{eq:DLF3}), and \emph{lower right panel}) together with their $\tau$-derivatives are all expressed in terms of the proper time $\tau$. The dashed blue lines in all plots represent the proper time $T_{\rm touch}$ at which the test particle reaches the critical hypersurface and it amounts to $T_{\rm touch}=2915M$.}
	\label{fig:Fig3}
\end{figure}

\begin{itemize}
\item \emph{The relative mechanical energy} of the test particle with respect to the critical hypersurface measured in the ZAMO frame is 
\begin{equation} \label{eq:LF1}
\mathbb{K}=\frac{m}{2}\left|\nu^2-\nu^2_{\rm crit}\right|+(A-M)\left(\frac{1}{r}-\frac{1}{r_{\rm crit}}\right),
\end{equation}
where $\nu_{\rm crit}(\theta)=[\cos\beta]_{r=r_{\rm crit}(\theta)}$, which includes as a particular case the velocity $\nu_{\rm eq}=[\cos\beta]_{r=r_{\rm crit}(\pi/2)}$ in the equatorial ring. Its derivative is 
\begin{equation}  \label{eq:DLF1}
\begin{aligned}
\dot{\mathbb{K}}&=m\ {\rm sgn}\left(\nu^2-\cos^2\beta\right)\left[\nu\frac{d\nu}{d\tau}-\cos\beta\frac{d (\cos\beta)}{d \tau}\right]-\frac{A-M}{r^2}\dot{r}.
\end{aligned}
\end{equation}
where ${\rm sgn}(x)$ is the signum function.
\item \emph{The angular momentum} of the test particle measured in the ZAMO frame is 
\begin{equation}  \label{eq:LF2}
\begin{aligned}
\mathbb{L}=m(r\nu\sin\psi\cos\alpha-r_{\rm crit}\nu_{\rm crit}).
\end{aligned}
\end{equation}
Its derivative is given by
\begin{equation}  \label{eq:DLF2}
\begin{aligned}
\dot{\mathbb{L}}&=m\ \left[-\dot{r}_{\rm crit}\nu_{\rm crit}-r_{\rm crit}\frac{d(\nu_{\rm crit})}{d\tau}+r\frac{d\nu}{d\tau}\cos\alpha\sin\psi+\nu(\dot{r}\cos\alpha\sin\psi\right.\\
&\left.-r\sin\alpha\sin\psi\ \dot{\alpha}+r\sin\alpha\cos\psi\ \dot{\psi})\right].
\end{aligned}
\end{equation}
\item \emph{The Rayleigh dissipation function} is (see Sec. \ref{sec:RF} for its derivation and meaning)
\begin{equation}
\mathbb{F}=\tilde{\sigma}\mathcal{I}^2\left[\lg\left(\frac{\mathbb{E_{\rm crit}}}{E_p}\right)-\lg\left(\frac{\mathbb{E}}{E_p}\right)\right],
\end{equation}
where $E_p$ is the photon energy and $\mathbb{E}\equiv E(U)$, defined as
\begin{equation}  \label{eq:LF3}
\begin{aligned}
\mathbb{E}&\equiv-k_\alpha U^\alpha=\gamma \frac{E_p}{N}(1+bN^\varphi)[1-\nu\sin\psi\cos(\alpha-\beta)].
\end{aligned}
\end{equation}
$\mathbb{E_{\rm crit}}$ is the energy $\mathbb{E}$ evaluated on the critical hypersurface, given by 
\begin{equation} 
\begin{aligned}
\mathbb{E_{\rm crit}}&=[\mathbb{E}]_{r=R_\star,\alpha=0,\pi,\psi=\pm\pi/2,\nu=\nu_{\rm crit}}=\frac{E_p|(\sin\beta)_{\rm crit}|}{N_{\rm crit}}(1+bN^\varphi_{\rm crit}).
\end{aligned}
\end{equation}
Its derivative is
\begin{equation}  \label{eq:DLF3}
\begin{aligned}
\dot{\mathbb{F}}&=\tilde{\sigma}\dot{(\mathcal{I}^2)}\left[\lg\left(\frac{\mathbb{E_{\rm crit}}}{E_p}\right)-\lg\left(\frac{\mathbb{E}}{E_p}\right)\right] +\tilde{\sigma}\mathcal{I}^2\left[\frac{\dot{\mathbb{E}}_{\rm crit}}{\mathbb{E_{\rm crit}}}-\frac{\dot{\mathbb{E}}}{\mathbb{E}}\right].
\end{aligned}
\end{equation}
\end{itemize}
In Fig. \ref{fig:Fig3} we calculate a test particle orbit in the equatorial plane reaching the critical hypersurface, and in the other panels we show the three proposed functions (i.e., $\mathbb{K},\ \mathbb{L},\ \mathbb{F}$) together with their derivatives (i.e., $\dot{\mathbb{K}},\ \dot{\mathbb{L}},\ \dot{\mathbb{F}}$), to graphically prove that they verify the three proprieties to be Lyapunov functions. 

It is important to note that the first two Lyapunov functions (energy and angular momentum) are written using the classical definition, and not the general relativistic expression, as instead it has been done with the third Lyapunov function. This is not in contradiction with the definition of Lyapunov function, rather they are very useful to carry out more easily the calculations. For example, even a mathematical function with no direct physical meaning with the system under study, but verifying the conditions to be a Lyapunov function is a good candidate to prove the stability of the critical hypersurfaces.

\section{Analytical form of the Rayleigh dissipation function}
\label{sec:RF}
We describe the energy formalism, which is the method used to derive the Rayleigh potential \cite{Defalco2019EF2}. The motion of the test particle occurs in $\mathcal{M}$, a simply connected domain (the region outside of the compact object including the event horizon). We denote with $T\mathcal{M}$ the tangent bundle of $\mathcal{M}$, whereas $T^*\mathcal{M}$ stands for the cotangent bundle over $\mathcal{M}$. Let $\boldsymbol{\omega}:T\mathcal{M}\rightarrow T^*\mathcal{M}$ be a smooth differential semi-basic one-form. Defined $\boldsymbol{X}=(t,r,\theta,\varphi)$ and $\boldsymbol{U}=(U^t,U^r,U^\theta,U^\varphi)$, the radiation force components (\ref{radforce}) are the components of the differential semi-basic one-form $\boldsymbol{\omega}(\boldsymbol{X},\boldsymbol{U})=F_{\rm (rad)}(\boldsymbol{X},\boldsymbol{U})^\alpha \boldsymbol{{\rm d}}X_\alpha$. We note that $\boldsymbol{\omega}$ is closed under the vertical exterior derivative $\boldsymbol{{\rm d^V}}$ if $\boldsymbol{{\rm d^V}}\boldsymbol{\omega} = 0$. The local expression of this operator is
\begin{equation} \label{eq:vertical_derivative1}
\boldsymbol{{\rm d^V}}F= \frac{\partial F}{\partial U_\alpha} \boldsymbol{{\rm d}}X_\alpha.
\end{equation}
For the Poincar\'e lemma (generalised to the vertical differentiation) the closure condition and the simply connected domain $\mathcal{M}$ guarantee that $\boldsymbol{\omega}$ is exact. Therefore, it exists a 0-form $V(\boldsymbol{X},\boldsymbol{U})\in \mathcal{C}^\infty(T\mathcal{M},\mathfrak{m})$, called primitive, such that $-\boldsymbol{{\rm d^V}} V=\boldsymbol{\omega}$. 

 Due to the non-linear dependence of the radiation force on the test particle velocity field, the semi-basic one-form turns out to be not exact \cite{Defalco2018}. However, the PR phenomenon exhibits the peculiar propriety according to which $\boldsymbol{\omega}(\boldsymbol{X},\boldsymbol{U})$ becomes exact through the introduction of the integrating factor $\mu =\left(E_{\rm p}/\mathbb{E}\right)^2$ \cite{Defalco2018}. Considering the energy $\mathbb{E}=-k_\beta U^\beta$ and substituting all the occurrences of $\mathbb{E}$ in $F_{\rm (rad)}(\boldsymbol{X},\boldsymbol{U})^\alpha$, see Eq. (\ref{radforce}), we obtain \cite{Defalco2019EF1,Defalco2019EF2} 
\begin{equation}
\mathbb{F}_{\rm (rad)}(\boldsymbol{X},\boldsymbol{U})^\alpha=-k^\alpha \mathbb{E}(\boldsymbol{X},\boldsymbol{U})+\mathbb{E}(\boldsymbol{X},\boldsymbol{U})^2 U^\alpha.
\end{equation}
Using the chain rule from the velocity to the energy derivative operator, we have 
\begin{equation} \label{eq:trader}
\frac{\partial\ (\ \cdot\ )}{\partial U_\alpha}=-k^\alpha\ \frac{\partial\ (\ \cdot \ )}{\partial \mathbb{E}}.
\end{equation}
Therefore, the $V$ function satisfies the usual primitive condition \cite{Defalco2019EF1,Defalco2019EF2}
\begin{equation} \label{eq:primitive_original}
\mu F_{\rm (rad)}^\alpha=k^\alpha\frac{\partial V}{\partial \mathbb{E}}.
\end{equation}
Such differential equation for $V$ contains the $k^\alpha$ factor, which represents an obstacle for the integration process. To get rid of this term, we can consider the scalar product of both members of Eq. (\ref{eq:primitive_original}) by $U_\alpha$, which permits to obtain a more manageable integral equation for $V$ \cite{Defalco2019EF1,Defalco2019EF2}, i.e., 
\begin{equation} \label{eq:pot_E}
V=-\int \left(\frac{\mu F^\alpha}{\mathbb{E}} \right){\rm d}\mathbb{E}+f(\boldsymbol{X},\boldsymbol{U}),
\end{equation}
where $f$ is constant with respect to $\mathbb{E}$, i.e., $\partial f/\partial \mathbb{E}=0$ and $V$ is still a function of the local coordinates $(\boldsymbol{X},\boldsymbol{U})$. Integrating Eq. (\ref{eq:pot_E}), the final result is \cite{Defalco2019EF1,Defalco2019EF2}
\begin{equation} \label{eq: Rayleigh_potential_final}
V=\tilde{\sigma}\mathcal{I}^2\left[\ln\left(\frac{\mathbb{E}}{E_{\rm p}}\right)+\frac{1}{2}\left(U_\alpha U^\alpha+1\right)\right].
\end{equation}
Equation (\ref{eq: Rayleigh_potential_final}) is consistent with the classical description \cite{Poynting1903,Robertson1937}. The PR effect configures as the first dissipative non-linear system in GR for which we know the analytical form of the Rayleigh potential. 

\subsection{Discussions of the results}
The energy function $\mathbb{E}$ and the chain rule both represent the fundamental aspects of the energy formalism, since they permit to simplify the demanding calculations for the $V$ primitive. We are able to substantially reduce the coordinates involved in the calculations, passing from $4$ initial parameters, represented by $\boldsymbol{U}$, to one only, i.e., the energy $\mathbb{E}$. In particular, the $f$ function occurring in Eq. (\ref{eq:pot_E}) embodies our ignorance about the analytic form of $V$ as a function of the local coordinates $(\boldsymbol{X},\boldsymbol{U})$. In some cases, as in our model, the $f$ function can be determined by applying the integration process for an exact differential one-form. Such method has also the peculiar propriety, that it is independent from the considered metric, permitting to be applied to generic metric theories of gravity, and for its generality and simplicity, it can be also applied in different physical and mathematical fields. 
\begin{figure}[t!]
	\centering
	\includegraphics[scale=0.46]{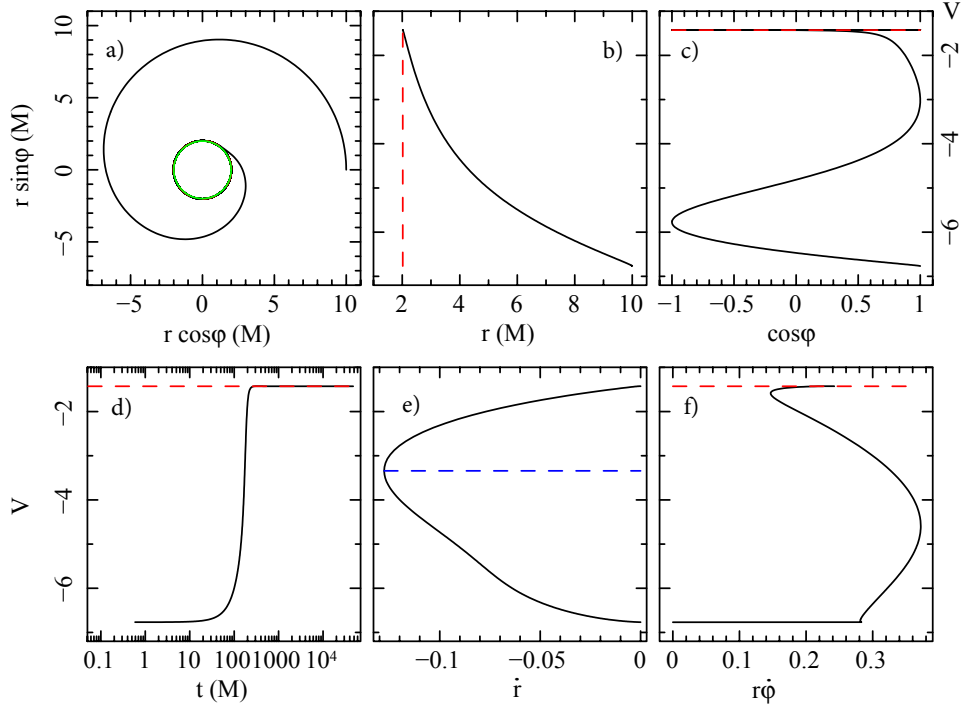}
	\caption{Test particle trajectory with the Rayleigh potential $V$ for mass $M=1$ and spin $a=0.1$, luminosity parameter $A=0.1$ and photon impact parameter $b=1$. The test particle moves in the spatial equatorial plane with initial position $(r_0,\varphi_0)=(10M,0)$ and velocity $(\nu_0,\alpha_0)=(\sqrt{1/10M},0)$. a) Test particle trajectory spiralling towards the BH and stopping on the critical radius (red dashed line) $r_{\rm (crit)}=2.02M$. The continuous green line is the event horizon radius $r^+_{\rm (EH)}=1.99M$. Rayleigh potential versus b) radial coordinate, c) azimuthal coordinate, d) time coordinate, e) radial velocity, and f) azimuthal velocity. The blue dashed line in panel e) marks the minimum value attained by the radial velocity, corresponding to $\dot{r}=-0.13$.}
	\label{fig:Fig4}
\end{figure}

The Rayleigh potential (\ref{eq:pot_E}) is a valuable tool to investigate the proprieties of the general relativistic PR effect and more in general the radiation processes in high-energy astrophysics. This potential contains a logarithm of the energy, which physically is interpreted as the absorbed energy from the test particle. Therefore, it represents a new class of functions, never explored and discovered in the literature, used to descrive the radiation absorption processes. Another important implication of the Rayleigh potential relies on the direct connection between theory and observations. In Fig. \ref{fig:Fig4} we show in panel $a)$ the test particle trajectory (what we can observe) and in panels $b)-f)$ the Rayleigh potential in terms of the coordinates $r,\varphi,t,\dot{r},\dot{\varphi}$, respectively (what comes from the theory). Therefore, observing the test particle motion, it is possible to theoretically reconstruct the Rayleigh function; viceversa new Rayleigh functions can be proposed to study then the dynamics and see what we should observe (see Ref. \cite{Defalco2019EF2}, for details).  

\section{Conclusions}
\label{sec:end}
In this work, we have presented the fully general relativistic treatment of the 3D PR effect in the Kerr geometry, which extends the previous works framed in the 2D equatorial plane of relativistic compact objects. The radiation field comes from a rigidly rotating spherical source around the compact object. The emitted photons are parametrized by two impact parameters $(b,q)$, where $b$ can be variable and $q$ depends on the value assumed by $b$ and the polar angle $\theta$, position occupied by the test particle in the 3D space. The resulting equations of motion represent a system of six coupled ordinary and highly nonlinear differential equations of first order. The motion of test particles is strongly affected by PR effect together with general relativistic effects. Such dynamical system admits the existence of critical hypersurfaces, regions where the gravitational attraction is balanced by the radiation forces. 

We have presented the method to prove the stability of the critical hypersurfaces by employing the Lyapunov functions. Such strategy permits to simplify the calculations and to catch important physical aspects of the PR effect. Three different Lyapunov functions have been proposed, all with a different and precise meaning. The first two are deduced by the definition of the PR effect, which removes energy and angular momentum from the test particle. The third example is less intuitive because it is based on the Rayleigh dissipation function, determined by the use of an integrating factor and the introduction of the energy formalism. 

Such method revealed to be very useful for two reasons: (1) a substantial reduction of the calculations from the $4$ variables (i.e., the velocities $\boldsymbol{U}$) to only one (i.e., the energy $\mathbb{E}$); (2) the obtained expression of the $V$ potential as a function of $\mathbb{E}$, suffices for the description of the dynamics, being very important whenever the evaluation of $f(\boldsymbol{X},\boldsymbol{U})$ turns out to be too laborious. In this way we have obtained for the first time an analytical expression of the Rayleigh potential in GR and we have discovered a new class of functions, represented by the logarithms, which physically describe the absorption processes in high-energy astrophysics.

As future projects, we plan to improve the actual theoretical assessments used to treat the radiation field in some ingenue aspects, like: the momemntum-transfer cross section will be not anymore constant, but it will depend on the angle and frequency of the incoming radiation field, the radiation field is not emitted anymore by a point-like source, but from a finite extended source. We would like also to apply this theoretical model to some astrophysical situations in accretion physics, like: accretion disk model, type-I X-ray burst, photospheric radius expansion. 

The new method to prove the stability of the critical hypersurfaces through Lyapunov functions can be easily applied to any possible extensions of the general relativistic PR effect model, naturally with the due modifications. Instead, the energy formalism opens up new frontiers in the study of the dissipative systems in metric theory of gravity and more broadly in other mathematical and physical research fields thanks to its general structure and versatile applicability. It permits to acquire more information on the mathematical structure and the physical meaning of the problem under study, because as discussed in Fig. \ref{fig:Fig4}, it is incredibly evident the profound connection between observations and theory.

\subsection*{Acknowledgment}
The author thanks the Silesian University in Opava and Gruppo Nazionale di Fisica Matematica of Istituto Nazionale di Alta Matematica for support. The results contained in the present paper have been partially presented at the summer school DOOMOSCHOOL 2019.

\end{document}